\documentclass[10pt,aps,pre,twocolumn,notitlepage,superscriptaddress,preprintnumbers]{revtex4-1}
\usepackage{amsmath,amssymb,amsfonts}
\usepackage{bm}
\usepackage{graphicx,color}
\usepackage{verbatim}
\usepackage{float}
\usepackage[caption = false]{subfig}
\usepackage{dcolumn}
\usepackage{soul}
\usepackage{natbib}
\usepackage{hyperref}
\usepackage{tikz}
\usetikzlibrary{hobby}
\usepackage{appendix}

\begin{document}
	\title{Quartic balance theory: Global minimum with imbalanced triangles}
	\author{A. Kargaran}
	\affiliation{Department of Physics, Shahid Beheshti University, G.C., Evin, Tehran 19839, Iran}
	\author{M. Ebrahimi}
	\affiliation{Department of Physics, Shahid Beheshti University, G.C., Evin, Tehran 19839, Iran}
	\author{M. Riazi}
	\affiliation{Department of Physics, Sharif University of Technology, P.O. Box 11365-9161, Tehran, Iran}
	\author{A. Hosseiny}
	\email{al\_hosseiny@sub.ac.ir}
	\affiliation{Department of Physics, Shahid Beheshti University, G.C., Evin, Tehran 19839, Iran}
	\author{G. R. Jafari}
	\email{g\_jafari@sbu.ac.ir}
	\affiliation{Department of Physics, Shahid Beheshti University, G.C., Evin, Tehran 19839, Iran}
	\affiliation{Department of Network and Data Science, Central European University, Nador u. 9, H-1051 Budapest, Hungary}
	\date{\today}
	\begin{abstract}
		Balance theory proposed by Heider for the first time modeled triplet interaction in a signed network, stating that relationships between two people, friendship or enmity, is dependent on a third person. The Hamiltonian of this model has an implicit assumption that all triads are independent, meaning that the type of each triad, being balanced or imbalanced,  determined apart from the state of other triads. This independence forces the network to have completely balanced final states. However, there exists evidence indicating that real networks are partially balanced raising the question of what is the mechanism preventing the system to be perfectly balanced. Our suggestion is to consider a quartic interaction which dissolves the triad's independence. We use the mean field method to study the thermal behavior of such systems where the temperature is a parameter that allows the stochastic behavior of agents. We show that under a certain temperature, the symmetry between balanced and imbalanced triads will spontaneously break and we have a discrete phase transition. As a consequence, stability arises where either similar balanced or imbalanced triads dominate, hence the system obtains two new imbalanced stable states. In this model, the critical temperature depends on the second power of the number of nodes, which was a linear dependence in thermal balance theory. Our simulations are in good agreement with the results obtained by the mean field method.
	\end{abstract}
	\maketitle
	\section{\label{sec:level1}Introduction} 
	{Thanks to the balance theory introduced by Heider \cite{heider1}\cite{heider2} we can go beyond describing a society only by its pairwise interactions. Heider balance theory (HBT) brilliantly considers triplet interactions between individuals and suggests that a tendency towards reducing psychological stresses is a key factor in the evolution of interrelationships. Later, Cartwright and Harary \cite{cartwright} used graph theory to extend the idea formerly proposed by Heider. Since then, HBT has been successfully applied to many branches of science including but not limited to studies of international relations \cite{hart,galam,bramson}, sociology  \cite{singh,szell,altafini,oloomi,sheykhali,hedayatifar,hassanibesheli,saeedian,fereshteh} and ecology \cite{saiz}. Moreover, many researches have expanded the theoretical aspects of HBT \cite{facchetti,estrada2,estrada3} in which one of the most controversial questions is the length of interaction. Here to properly address this question, we have theoretically expanded HBT and propose the possible idea of Quartic Balance Theory (QBT) by taking into account not only triplets but the quartic interactions as well.  
	
  	According to HBT, a community is modeled by a graph in which individuals are nodes and relationships between individuals are edges that can be positive (negative) representing friendship (enmity). Based on this framework, any triplet interaction is one of the four following triads, i.e., $[+ + +]$, $[+ - -]$, $[- - -]$ and $[+ - +]$. While the first two triads that have an even number of negative links are called balanced, the other two with an odd number of negative links are known to be imbalanced. This definition is based on the familiar idea that a friend of my friend is my friend as well, $[+ + +]$, and an enemy of my enemy is my friend, $[+ - -]$. However, in the case of imbalanced triads either there are three enemies interacting, $[- - -]$, or I have two friends that are enemies, $[+ - +]$. These imbalanced triads have a tendency to become balanced, in order to reduce the overall stress. The structural balance is reached when all the triads in the network become balanced. In this case two structures can emerge either a heaven or a bipolar. In heaven all triplets are of the $[+ + +]$ type, yet in bipolar the network is divided into two opposing subsets with positive links inside each subset and negative links between them. The bipolar was identified by Cartwright and Harary \cite{cartwright} through analyzing all interactions of different lengths. Accordingly, a network is balanced if and only if all paths joining the same pair of points have the same sign. Davis \cite{davis} also investigated bipolars with more than two balanced subsets.
  	
	Furthermore, many physicists considered social networks as complex systems and studied their properties \cite{cimini}. Marvel \textit{et al.} \cite{marvel1} defined a Hamiltonian for social balance and used statistical mechanics to investigate the local minima properties of the energy landscape, that is, the so-called jammed state. Belaza \textit{et al.} considered a more complex Hamiltonian \cite{belaza1} and brought into account the role of inactive links \cite{belaza2}. Antal \textit{et al.} studied discrete-time dynamics of how imbalanced triads turn into balanced and examined the evolution of social networks \cite{antal1}\cite{antal2}. They noted that on a fully connected network different stationary states can emerge. Continuous-time dynamics has been also discussed by Ku\l akowski \textit{et al.} through defining a set of differential equations \cite{kulakowski}. Afterwards, Marvel \textit{et al.} proved that the solutions of these differential equations are heaven and bipolar independent of the initial condition \cite{marvel2}. Taking a different approach, the stable states of the multiagent-based model is discussed \cite{caram1}\cite{caram2}\cite{sonubi}\cite{simin}, where the friendship (enmity) interactions between agents were described by generalized the Verhulst-Lotka-Volterra model.
	
	Along with these theoretical researches, HBT has also been put into practice to analyze real-world data. Specifically, HBT has been applied to study whether large-scale social networks are structurally balanced. Leskovec \textit{et al.} \cite{leskovec} stated that real-world networks are not actually fully balanced as we have expected. Such observations led scientists to reconsider the notion of balance and start redefining it as a degree instead of a dichotomous concept. Thus, different studies defined various measures for the degree of balance \cite{estrada2}\cite{kirkely}. For example, Facchetti \textit{et al.} \cite{facchetti} identified a measure of balance using the spin glass model, based on which the real-world networks found to be extremely balanced. However, surprisingly according to Estrada and Benzi \cite{estrada2}\cite{estrada3} that have defined a walk-based measure of balance, online networks are actually poorly balanced. The results of these and other studies, although very interesting and promising, have tremendous inconsistencies with each other. One of the many possible reasons for these discordant results is that different researchers have considered different lengths of interaction leading to different outcomes. For example, Estrada and Benzi \cite{estrada2} considered all cycles with different lengths while giving more weight to small ones. However, the length of interaction in the model by Facchetti \textit{et al.} \cite{facchetti} is one. Moreover, the length of interaction in many studies is three, that is, triplet interactions. These variations call for more in detailed theoretical studies regarding the length of interaction.
	
	Thus, in this study we move one step beyond triplets towards higher order interactions and bring quartic interactions into account. We introduce a Hamiltonian that treats both balanced and imbalanced triads equally depending on the status of neighbors (Sec \ref{model}). This is crucial because from Heider’s outlook, not only balanced but also imbalanced triads can be present in the stable states of a model \cite{heider2}.
	Additionally, this study makes it possible to observe the very impact that units of four entities have on the final degree of balance, which is of concern for two reasons. First, the basic social units as Heider defined are not actually limited to triplets, and as he discussed the interrelationships between three members can be influenced by the fourth person in the group \cite{cartwright} \cite{horowitz}. Second, by considering only quartic interactions at a time instead of taking into account different lengths altogether, we can step by step move beyond triplets towards higher order interactions and analyze each step thoroughly. Last but not least, we consider temperature as a measure of uncertainty which happens randomly in societies. The mathematical framework of our analytic approach is the statistical physics, especially \textit{exponential random graphs} \cite{holland,strauss1,strauss2,wasserman,anderson,snijders1,robins,cranmer,snijders2,newman1,newman2,newman3}, that we use to find mean value quantities. Eventually, we confirm our analytical solutions via simulation in Sec \ref{analysis}.
	\section{Model}\label{model}
	Following all of the previous statements, we consider a simple pairwise interaction term between triads with a common edge, as
	\begin{equation}\label{Hamiltonian}
		\begin{aligned}
			\mathcal{H}(G) & =-\sum_{i<j<k<\ell}\Delta_{ijk}\Delta_{ij\ell}\\
			& =-\sum_{i<j<k<\ell}\sigma_{ij}\sigma_{jk}\sigma_{ki}\sigma_{ij}\sigma_{j\ell}\sigma_{\ell i} \\
			& =-\sum_{i<j<k<\ell}\sigma_{jk}\sigma_{ki}\sigma_{j\ell}\sigma_{\ell i}=-s(G),
		\end{aligned}
	\end{equation}
	where, $\Delta_{ijk}$ represents a triad shaped by  $i, j, k$ nodes and $ \sigma_{ji}=\sigma_{ij}$ is an element of the adjacency matrix which connects node $ i $ to $ j $. The value of edges in adjacency matrix will be $ \pm 1 $ which defines the friendship or enmity relation between two nodes; so $\sigma_{ij}^2=1$ and double appearance of $\sigma_{ij}$ cancels its effect. In the above equations the number of squares, i.e., $ s(G) $, is an important parameter for the specific graph configuration $ G $.
	
	To understand why our Hamiltonian will result in local states with low energy, consider a fully connected network with four nodes. The Hamiltonian of this network consists of six terms, as there are six edges that can be counted as a common edge. Consider two triads in this network that have a common link. We can label each triad with a sign which is the product of its edges' signs. This triad's sign would have counted for balanced and imbalanced states in structural balance theory: negative sign for the imbalanced and positive sign for balanced. There are three structurally different possibilities for the combination of these two triads: $ ++ $, $ +- $, $ -- $. The first combination, in which both triads having positive signs, holds a model for four people having the same ideology towards an issue. In structural balance, this community has the lowest energy, being made of two balanced triads. The same holds in our model: the product of four edges will result in the minimum energy. The second combination has higher energy than the previous one, both in structural balance theory and in our model. However, the difference arises in the third combination: both triads being structurally imbalanced, will result in a state with lower energy in our model. Our model favors the configurations in which neighbors have the same structure by assigning lower energy to them. This Hamiltonian will result in lower energy for local subcommunities that have triads with the same structure.
	
	There exist some similarities between our model and the Ising model \cite{ising}. The Ising's Hamiltonian considers the pairwise interactions between spin sites. This Hamiltonian is in a sense similar to the Hamiltonian of our model (\ref{Hamiltonian}) if we consider each triad similar to a spin site. However, there exist more degrees of freedom in our model, for each spin site only has two possible configurations, but each triad has eight. We prefer to look at our Hamiltonian in a sense of ''square term,'' rather than the triad form.
	
	We consider the temperature in our model as a measure of randomness and use exponential random graph to obtain the probability distribution function \cite{book}. This function is actually the Boltzmann probability in canonical ensemble, $\mathcal{P}(G)\propto e^{-\beta \mathcal{H}(G)} $, where $\beta=1/T$.
	\section{Analysis}\label{analysis}
	\subsection{Mean-field solution}
	For the beginning we want to calculate the mean value of edges like $ \langle\sigma_{jk}\rangle $ over all configurations of our network. We rewrite our Hamiltonian as $ \mathcal{H}=\mathcal{H}'+\mathcal{H}_{jk} $ by separating all the terms containing $\sigma_{jk}$:
	\begin{equation}
		-\mathcal{H}_{jk}=\sigma_{jk}\sum_{i\neq j,k}\sum_{\ell\neq j,k}\sigma_{ki}\sigma_{j\ell}\sigma_{\ell i}.
	\end{equation}
	We can infer from statistical mechanics
	\begin{equation}
		\langle\sigma_{jk}\rangle=\sum_{G}\sigma_{jk}\mathcal{P}(G),
	\end{equation}  
	where $ \mathcal{P}(G)=e^{-\beta\mathcal{H}(G)}/\mathcal{Z} $ is Boltzmann probability and $\mathcal{Z}=\sum_{G}e^{-\beta\mathcal{H}(G)} $ is the partition function, and we have
	\begin{figure}[t]
		\centering
		\includegraphics[scale=1]{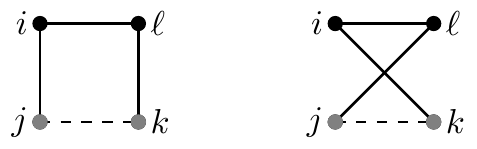}
		\caption{$ \sigma_{jk} $ involves two configurations with two selected nodes, as shown here.} \label{fig:conf}
	\end{figure}
	\begin{equation}\label{single-edge}
		\begin{aligned}
			&\langle\sigma_{jk}\rangle=\frac{1}{\mathcal{Z}}\sum_{\{\sigma\neq\sigma_{jk}\}}e^{-\beta\mathcal{H}'}\sum_{\sigma_{jk}=\pm 1}\sigma_{jk}e^{-\beta\mathcal{H}_{jk}} \\
			&=\frac{\sum_{\{\sigma\neq\sigma_{jk}\}}e^{-\beta\mathcal{H}'}\left[e^{-\beta\mathcal{H}_{jk}(\sigma_{jk}=+1)}-e^{-\beta\mathcal{H}_{jk}(\sigma_{jk}=-1)}\right]}{\sum_{\{\sigma\neq\sigma_{jk}\}}e^{-\beta\mathcal{H}'}\left[e^{-\beta\mathcal{H}_{jk}(\sigma_{jk}=+1)}+e^{-\beta\mathcal{H}_{jk}(\sigma_{jk}=-1)}\right]}\\
			&=\frac{\left\langle e^{-\beta\mathcal{H}_{jk}(\sigma_{jk}=+1)}-e^{-\beta\mathcal{H}_{jk}(\sigma_{jk}=-1)}\right\rangle_{G'}}{\left\langle e^{-\beta\mathcal{H}_{jk}(\sigma_{jk}=+1)}+e^{-\beta\mathcal{H}_{jk}(\sigma_{jk}=-1)}\right\rangle_{G'}}.
			\\
		\end{aligned}
	\end{equation}
	By $\langle\cdots\rangle_{G'}$, we mean the average over all graph configurations that does not contain $ \sigma_{jk} $. Now we can expand the above fraction and estimate the higher order product using mean-field approximation. In this method, we approximate edges' variables with their averages and the correlation between these variables simply become the product of their averages. For example, if we label some edges' variable $ A $, we have $ \langle AA\rangle\approx \langle A\rangle^2 $. Equipped with this method, we can approximate the above quantity; if we name our edges' variables as $  o\equiv\langle\sigma_{ki}\sigma_{j\ell}\sigma_{\ell i}\rangle $ and $ p\equiv\langle\sigma_{jk}\rangle $, we have
	\begin{equation}\label{meanofedges}
		p=\tanh[\beta(n-2)(n-3)o].
	\end{equation}
	The coefficient in the above equation is the number of all possible squares which contain $ \sigma_{jk} $ and is equal to $2\times\binom{n-2}{2} $. The factor of 2 comes from the two possible configurations with two selected nodes, Fig. \ref{fig:conf}.
	\begin{widetext} 
	Now let us calculate the following mean quantities with the same method: $ q\equiv\langle\sigma_{jk}\sigma_{ki}\rangle $, which is the mean value of two edges sharing a node or as we call it, the mean of two stars; $ o\equiv\langle\sigma_{jk}\sigma_{ki}\sigma_{j\ell}\rangle $, the mean of open squares; $r\equiv \langle\sigma_{ij}\sigma_{jk}\sigma_{ki}\rangle $, the mean of triangles; and $ s\equiv\langle\sigma_{jk}\sigma_{ki}\sigma_{j\ell}\sigma_{\ell i}\rangle $, the mean of squares. Rewriting our Hamiltonian for calculating the mean of two stars as $ \mathcal{H}=\mathcal{H}'+ \mathcal{H}_{\vee}$ with $\mathcal{H}_{\vee}$ being
	\begin{equation}
		-\mathcal{H}_{\vee}=\sigma_{jk}\sigma_{ki}\sum_{\ell\neq i,j,k}\sigma_{j\ell}\sigma_{\ell i}+\sigma_{jk}\sum_{\mu\neq i,j,k}\sum_{\ell\neq i,j,k}\sigma_{k\mu}\sigma_{j\ell}\sigma_{\ell\mu}
		+\sigma_{ki}\sum_{\mu\neq i,j,k}\sum_{\ell\neq i,j,k}\sigma_{\mu k}\sigma_{\mu \ell}\sigma_{\ell i}
	\end{equation}
	and $ \mathcal{H}' $ is the remaining terms. Similar to above, we have
		\begin{equation}\label{meanoftwostars}
			\begin{aligned}
				\langle\sigma_{jk}\sigma_{ki}\rangle&=\sum_{G}\sigma_{jk}\sigma_{ki}\mathcal{P}
				(G)\\
				&=\frac{\left\langle e^{-\beta\mathcal{H}_{\vee}(\sigma_{jk}=1,\sigma_{ki}=1)}-e^{-\beta\mathcal{H}_{\vee}(\sigma_{jk}=-1,\sigma_{ki}=1)}-e^{-\beta\mathcal{H}_{\vee}(\sigma_{jk}=1,\sigma_{ki}=-1)}+e^{-\beta\mathcal{H}_{\vee}(\sigma_{jk}=-1,\sigma_{ki}=-1)}\right\rangle_{G'}}{ \left\langle e^{-\beta\mathcal{H}_{\vee}(\sigma_{jk}=1,\sigma_{ki}=1)}+e^{-\beta\mathcal{H}_{\vee}(\sigma_{jk}=-1,\sigma_{ki}=1)}+e^{-\beta\mathcal{H}_{\vee}(\sigma_{jk}=1,\sigma_{ki}=-1)}+e^{-\beta\mathcal{H}_{\vee}(\sigma_{jk}=-1,\sigma_{ki}=-1)}\right\rangle_{G'}},
			\end{aligned}
		\end{equation}
		where $ G' $ is all graph configurations that do not contain $ \sigma_{jk}$ and $\sigma_{ki} $. The mean-field approximation for above is
		\begin{equation}
			q=\frac{e^{2\beta(n-3)(n-4)o+\beta(n-3)q}-2\,e^{-\beta(n-3)q}+e^{-2\beta(n-3)(n-4)o+\beta(n-3)q}}{e^{2\beta(n-3)(n-4)o+\beta(n-3)q}+2\,e^{-\beta(n-3)q}+e^{-2\beta(n-3)(n-4)o+\beta(n-3)q}}.
		\end{equation} 
	Similarly, equations for mean open squares, triangles, and squares can be derived. The mean values of two stars and open squares are fundamental, which means by knowing these two, all mean quantities like the mean values of triangles and squares, will be calculated. In the Appendix an equation for the mean open value of squares is derived in more detail, and equations of the mean values of triangles and squares are also shown.
	\end{widetext}
	By substituting (\ref{meanofedges}) in (\ref{opensquares}) and using (\ref{meanoftwostars}) we can write self-consistency equations as 
	\begin{equation}\label{order-parameter-equations}
		\begin{aligned}
			q&= f(q,\,o\,;\;\beta,\, n),\\
			o&= g(q,\,o\,;\;\beta,\, n).
		\end{aligned}
	\end{equation}
	In Fig.~\ref{fig:Root-Intersections} we plot numeric solutions for  each equation separately on the $q$-$o $ plane (in their allowed domains $-1\le q\le 1$, $-1\le o\le 1$). The intersections of curves are our simultaneous solutions for both equations. Right figures show the number of intersections under and above the critical temperature. At $T=1050$ we have five intersections. If the temperature is higher than $T_c$ we have just one intersection, which is our trivial solution. The left diagram depicts the critical temperature, $T_c\approx 1062$, in which we have one intersection and two tangent curves. Depending on the temperature, we have one $(T>T_c)$, three $(T=T_c)$, or five solutions $(T<T_c)$. The number of solutions changes abruptly and it is the classical phenomenology of discrete phase transition.
	\begin{figure}[t]
		\centering
		\includegraphics[scale=0.2]{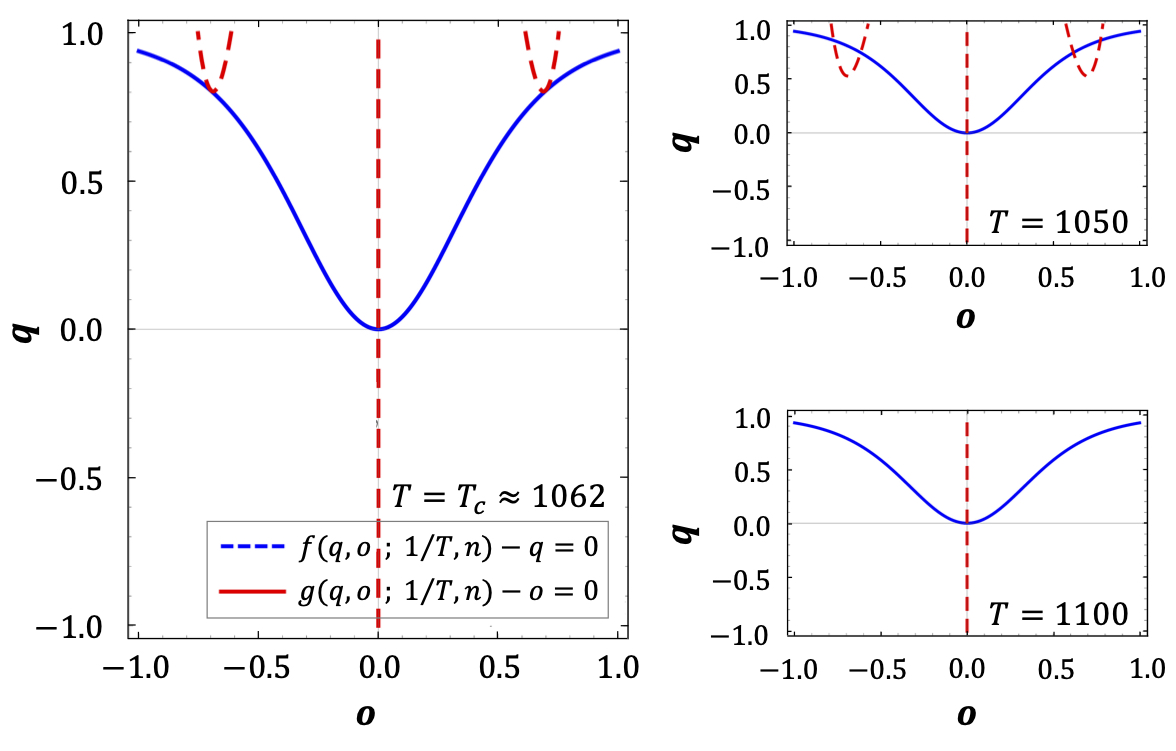}
		\caption{(Color online) Behavior of simultaneous solutions for (\ref{order-parameter-equations}) in different temperatures. The critical temperature is  $T_{c} \approx 1062$. There are five different solutions below (top right: $T=1050$) and one above the critical temperature (bottom right: $T=1100$). The number of nodes is $50$.}
		\label{fig:Root-Intersections}
	\end{figure}

	To discuss the stability of solutions or fixed points, we define a two-dimensional field with the following components:
	\begin{equation}
		\begin{aligned}
			u_q&\equiv f(q,\,o\,;\;\beta,\, n)-q,\\
			u_o&\equiv g(q,\,o\,;\;\beta,\, n)-o.\\
		\end{aligned}
	\end{equation}
	In Fig.~\ref{fig:q-o-rg} we show the vector field of the above quantities with (\ref{order-parameter-equations})'s solution for different temperatures. This plot shows the dynamic which $q$ and $o$ evolve to their final state. Below critical temperature $T<T_c$ we have five solutions, three of which are stable or attractive fixed points (blue circles) and the other two are unstable or repulsive fixed points (red triangles).    
	We can check the stability of these fixed points by considering a point in the $q$-$o $ plane very close to the specified fixed point $ (q^*,o^*) $; we have
	\begin{figure}[t]
		\centering
		\includegraphics[scale=0.18]{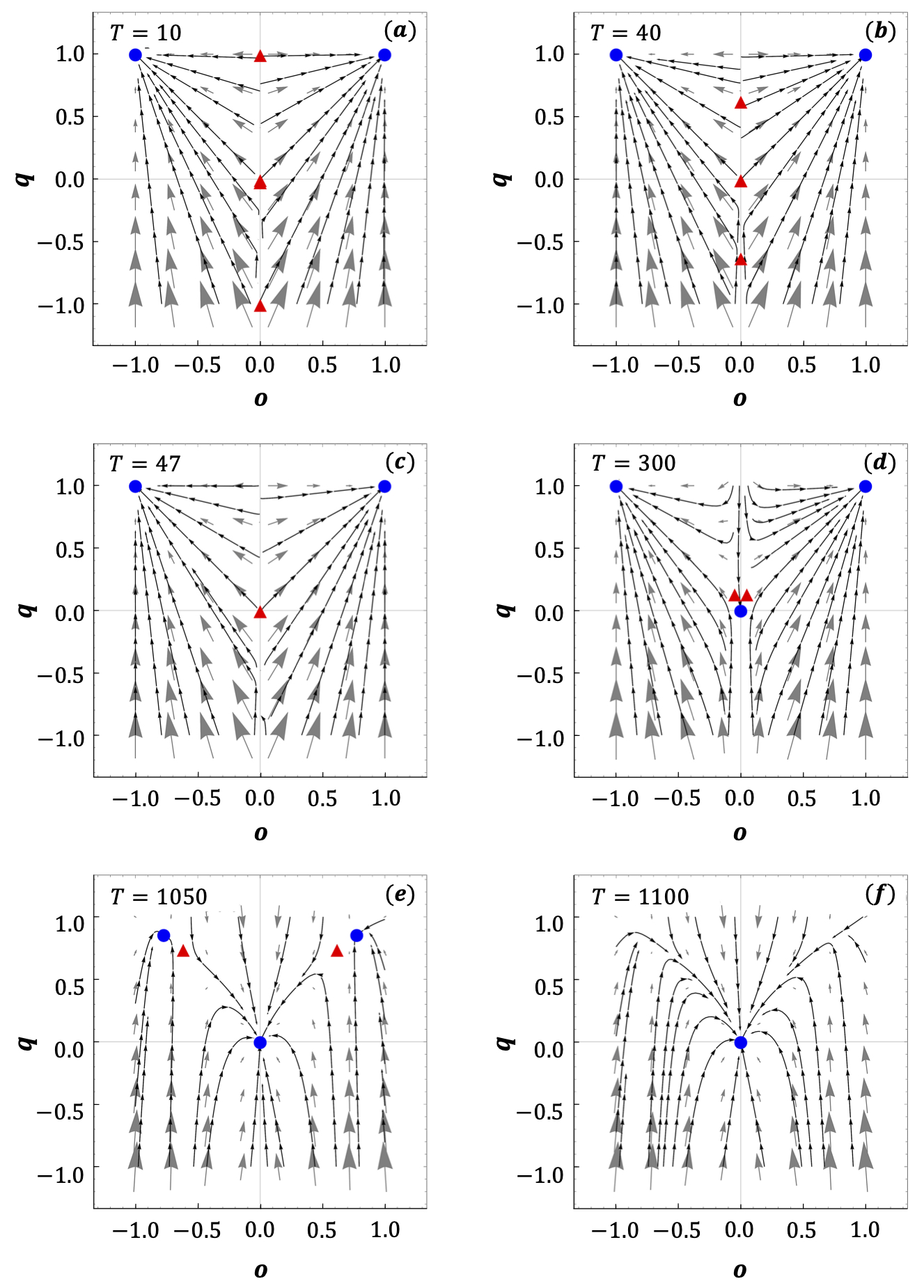}
		\caption{Stable (blue circles) and unstable (red triangles) fixed points for different temperatures in the $q$-$o $ plane. By increasing temperature, unstable fixed points move so close toward the trivial one in the origin [a, b, c], and they get away from it [d, e]. When the temperature is bigger than $T_c$ we have one stable fixed point (f). The number of nodes is 50. }
		\label{fig:q-o-rg}
	\end{figure}
	\begin{equation}\label{linear}
		\begin{aligned}
			q^*+\delta q'&= f(q^*+\delta q,\,o^*+\delta o\,;\;\beta,\, n),\\
			o^*+\delta o'&= g(q^*+\delta q,\,o^*+\delta o\,;\;\beta,\, n).
		\end{aligned}
	\end{equation}
	The Taylor expansion of the first equation will be
	\begin{equation}
		\begin{aligned}
			q^*+\delta q'&\approx f(q^*,o^*\,;\;\beta,\, n)\,\\
			& +\, \frac{\partial f}{\partial q}\Big |_{\substack{q=q^*\\o=o^*}}\delta q \, +\, \frac{\partial f}{\partial o}\Big |_{\substack{q=q^*\\o=o^*}}\delta o.
		\end{aligned}
	\end{equation}
	We consider the above expansion only in linear regime and the matrix form is
	\begin{equation}
		\left(\begin{array}{c} \delta q'\\ \delta o' \end{array}\right)  = \mathcal{J}\left(\begin{array}{c} \delta q \\ \delta o \end{array}\right), 
	\end{equation}
	where $\mathcal{J}$ is the Jacobian matrix
	\begin{equation}
		\mathcal{J}=
		\left(\begin{array}{cc} \partial f/\partial q & \partial f/\partial o\\ \partial g/\partial q & \partial g/\partial o \end{array}\right)_{\substack{q=q^*\\o=o^*}}.
	\end{equation}
	We can diagonalize the Jacobian matrix as
	\begin{equation}\label{digonal-form}
		\left(\begin{array}{c} \delta_{d}q' \\ \delta_{d}o' \end{array}\right)  =
		\left(\begin{array}{cc}\lambda_1 & 0\\ 0 & \lambda_2 \end{array}\right)\left(\begin{array}{c} \delta_{d} q \\ \delta_{d} o \end{array}\right).
	\end{equation}
	The magnitude of eigenvalues of the above equation can distinguish the stability of the fixed points. The only condition for the stable fixed point is $|\lambda_i|<1$ because the right hand side of (\ref{digonal-form}) becomes smaller by each iteration. At least one of the eigenvalues of the red triangles in  Fig.~\ref{fig:q-o-rg} is bigger than one which makes these fixed points unstable.
	
	We have three stable fixed points in a wide range of temperatures. The trivial solution $ (q^{*}=0,\;o^{*}=0) $ for (\ref{order-parameter-equations}) corresponds to a random or bipolar network, which means that the mean value of two stars and open squares is zero. This fixed point is unstable in low temperatures $ (T<50) $ and it becomes stable when the temperature increases. At low temperatures, the unstable fixed point is very close to the trivial solution [Fig.~\ref{fig:q-o-rg}(c)].
	
	We have two completely balanced states corresponding to heaven $ (q^*=+1,\, o^*=+1) $ and hell $ (q^*=+1,\, o^*=-1) $. In heaven's fixed point, all links are positive, which means that there is no hostility in the network. In hell's fixed point, all links are negative which means that all nodes are enemies to each other. These fixed points move when temperature increases and suddenly disappear when the temperature is higher than the critical temperature. 	
	\begin{figure}[t]
		\centering
		\includegraphics[scale=0.2]{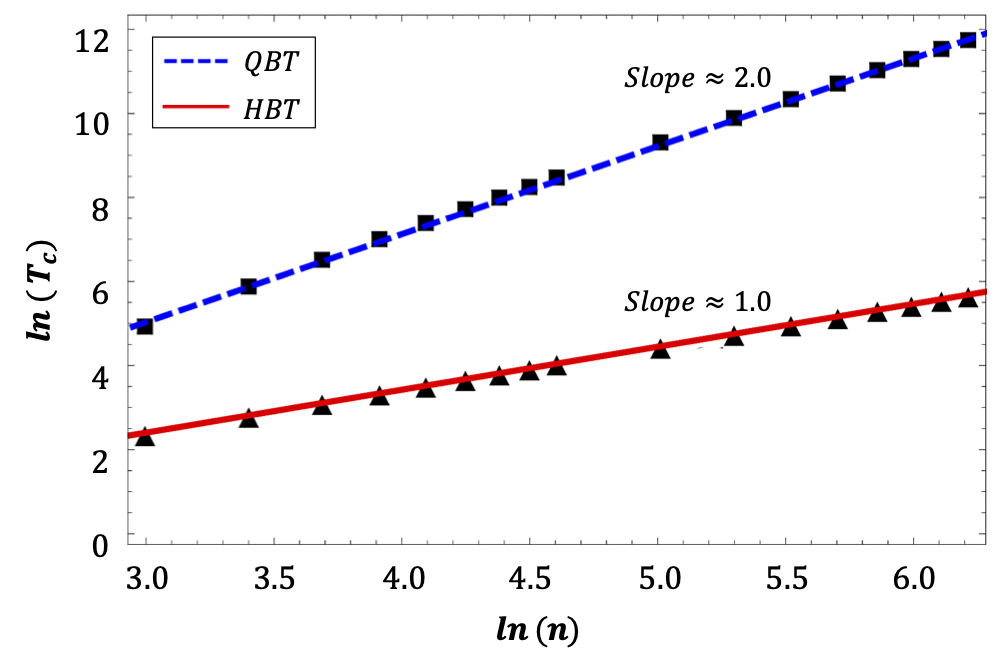}
		\caption{Behavior of critical temperature for the presented model (dashed blue line) and thermal HBT (solid red line) as a function of the number of nodes. Lines are fitted to data points of both models.}
		\label{fig:Tc-n}
	\end{figure}
	\begin{figure}
		\centering
		\includegraphics[scale=0.23]{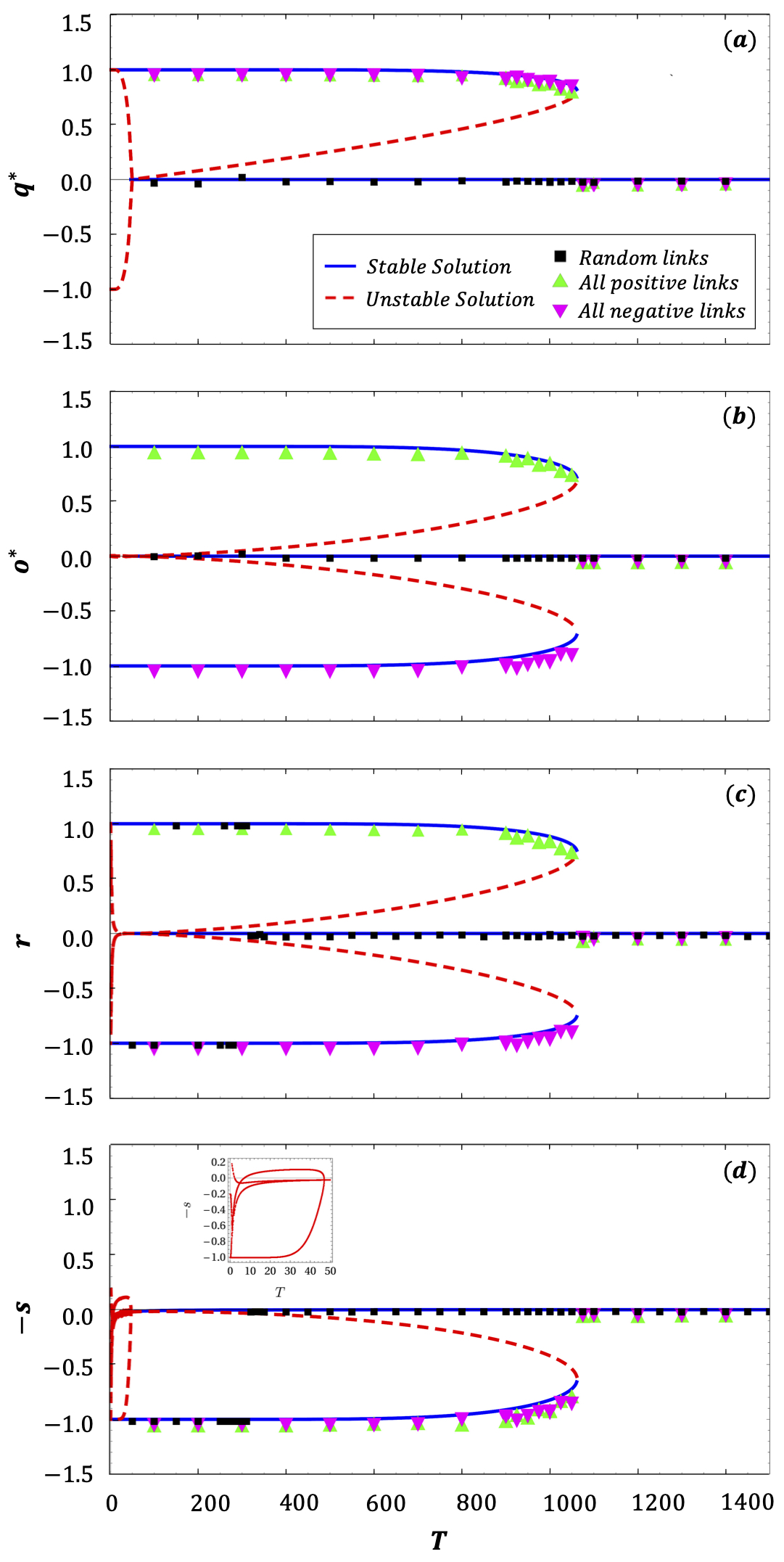}
		\caption{(Color online) Comparison of our analytic solutions (red dashed and solid blue lines) and Monte Carlo simulations (triangles and squares) for different initial network configurations, for $o$, $q$, $r$, and $s$. In (c) and (d) for $50<T<300$ unstable fixed points are so close to a stable fixed point and the system is pushed to ordered phases. Size of the network is 50.}
	\label{fig:q-o-r-s}
	\end{figure}
	In Fig.~\ref{fig:Tc-n} we compare the dependency of the critical temperature on the size of the network as in the presented model to that in the thermal balance theory. In quartic balance theory critical temperature changes as the square of size [$ T_c(n)\approx n^2 $] and in thermal balance theory it changes linearly [$ T_c(n)\approx n $].
	\subsection{Simulations}
	In our simulations, we work with a fully connected network with $ n $ nodes. We thermalize our system with a given temperature by the Monte Carlo method. In this method, we randomly pick an edge and compute the energy difference with the configuration where it is flipped. We accept this new configuration if this difference is negative. Also,  if this difference is positive, we accept the new configuration with the Boltzmann's probability.
	
	In Fig.~ \ref{fig:q-o-r-s} we compare our analytic solutions with simulations. Our theory gives us stable and unstable solutions and we can not see the unstable one (red dashed line) in simulation. We start our simulations with different initial configurations to find how our network changes when the randomness increases. 	
	\begin{figure}[t]
		\centering
		\includegraphics[scale=0.12]{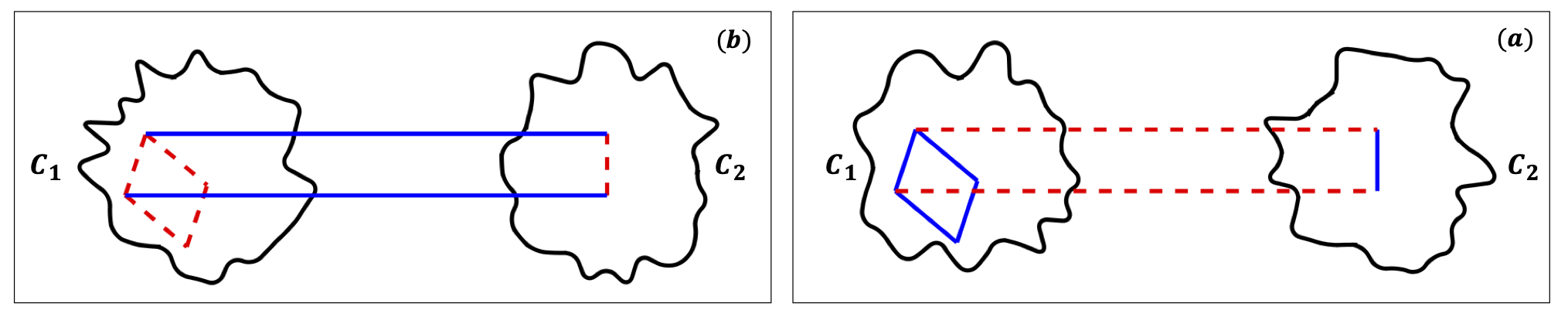}
		\caption{Ordered final networks with initial random configurations. In our model there are two possible bipolar configurations. (a) Two cliques ($ C_1 $ and $ C_2 $) with friendship links (solid line) inside and enmity links in between. (b) Two cliques with enmity links (dashed line) inside and friendship links in between.}
		\label{fig:random-bipolar}
	\end{figure}
	\begin{figure}[b]
		\centering
		\includegraphics[scale=0.17]{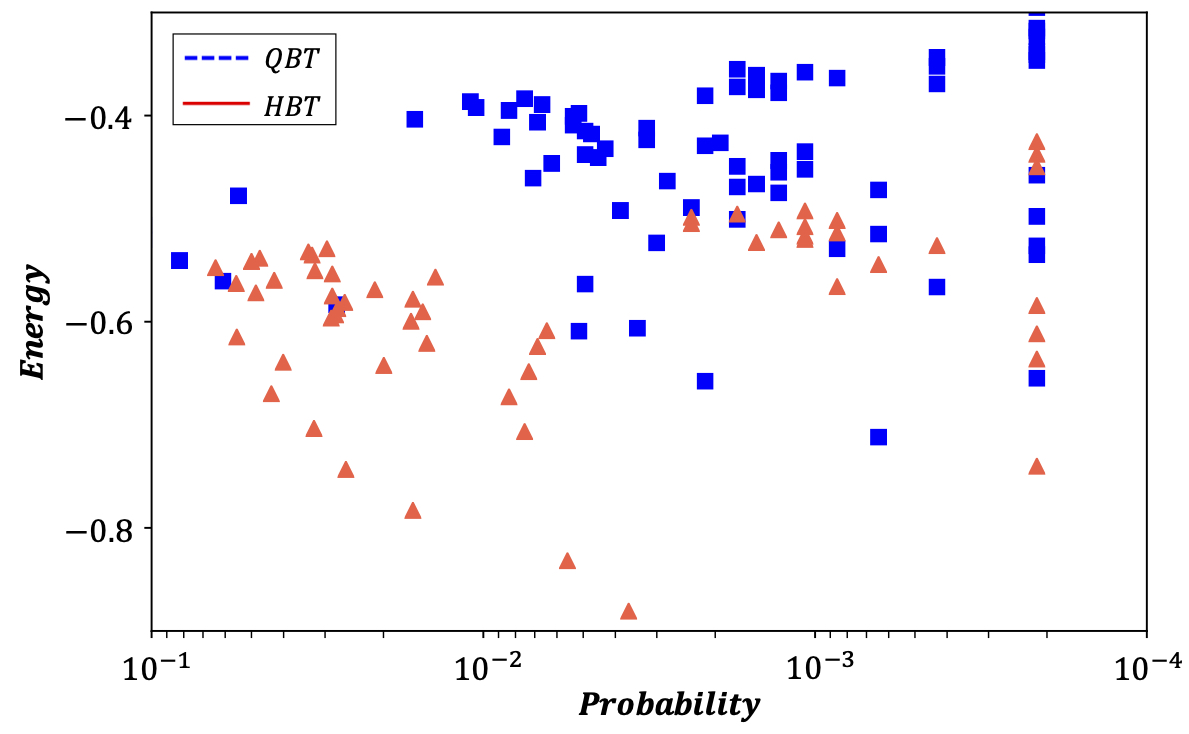}
		\caption{A comparison between the probability density of jammed states in QBT and HBT. The number of nodes is 50.}
		\label{fig:jammed}
	\end{figure}
	In random initial configurations, all kinds of balanced and imbalanced squares exist, which means that the total energy of our system is near zero ($ s\approx 0 $). It is interesting that with this initial condition and in a temperature range of $ 50<T<300 $ we expect from our theory that the trivial solution is stable and the system shall remain there forever; however, our unstable solution is so close to the stable one, which pushes our final state to the bipolar phases [Fig.~\ref{fig:q-o-r-s}(c)]. Bipolar networks have small mean values of two stars and open squares but the mean value of the triangles is close to one. In Fig.~\ref{fig:random-bipolar} we have plotted the bipolar networks of our model schematically.
	
	In Monte Carlo simulation, the final state of the network can be trapped in the minimum local energy. This occurs when the system's temperature is zero, to which the literature has referred as the jammed states. In Fig.~\ref{fig:jammed}, the probability of jammed states of the presented model is compared with HBT. These distributions are generated using $ 5\times 10^{3} $ jammed states for both models. As can be seen the QBT's jammed states are more likely to occur at higher energies compared to HBT. Higher energies in jammed states lead to more structural complexity \cite{marvel2}.
	\section{CONCLUSIONS}   
	We have theoretically studied the idea of expanding HBT based on Heider’s perspective that states, the tension which comes from imbalanced situations is not always to be dismissed, but to be embraced \cite{heider2}. The proposed model obtains the following results through a mean-field method: 
	
	(1) The global minimum of this model has four stable phases: heaven, hell, and two bipolars. Two of these phases include imbalance triads, that is, hell where all triads are of 
	$[- - -]$ type and one of the bipolars that consists of 
	$[- - -]$ and $[+ - +]$.
	
	(2) There is a phase transition where below a specific temperature a symmetry between balanced and imbalanced triads is broken, which is well consistent with the thermal HBT in a way that it is also discrete.
	
	(3) The critical temperature of QBT is proportional to the square of the network’s size. This is while, in thermal HBT the critical temperature changes linearly with size.
	
	(4) Our Monte Carlo simulations confirm the prediction of the discrete phase transition and the quantitative value of the critical temperature. 
	
	(5) As results of simulations indicate, in low temperatures a trivial solution is not stable anymore, which can be due to the narrowness of the basin of attraction. 
	
	(6) The probability distribution of the jammed states has shifted to a higher level of energy, compared to that in HBT.	
	
	These results are rooted in the dynamics that for triads the decision to change from balanced to imbalanced, is determined according to all triads that share a common link. Thus, in QBT the presence of imbalanced triads no longer contradicts with a network being structurally stable. It needs to be clearly stated that, the proposed model needs to be further examined on real signed networks to investigate its suitability and accuracy. Here we have presented the model theoretically, which needs to be explored in the future to assess its rightness.
	\begin{acknowledgments}                
	A.K. would like to express his appreciation to M. J. Kazemi, A. Banihashemi, H. Hashamipour, and B. Askari for constructive discussions. His thanks are extended to Z. Moradimanesh as well for comments that improved the manuscript.
	\end{acknowledgments}                
	\appendix
	\begin{appendices}
		\section{Calculation of The Mean Values of Open Squares}
		\label{appendix:mean quantity}
		In this part we want to calculate the mean value of open squares. As previously, we write the Hamiltonian as $ \mathcal{H}=\mathcal{H}'+ \mathcal{H}_{\sqcup}$, where
		\begin{equation}
			\begin{aligned}
				-\mathcal{H}_{\sqcup}&=\sigma_{jk}\sum_{\mu\neq i,j,k,\ell}\sum_{\nu\neq i,j,k,\ell}\sigma_{k\mu}\sigma_{j\nu}\sigma_{\nu\mu} \\
				& + \sigma_{ki}\sum_{\mu\neq i,j,k,\ell}\sum_{\nu\neq i,j,k,\ell}\sigma_{\mu k }\sigma_{\mu \nu}\sigma_{\nu i}\\
				& + \sigma_{j\ell}\sum_{\mu\neq i,j,k,\ell}\sum_{\nu\neq i,j,k,\ell}\sigma_{j\mu}\sigma_{\mu\nu}\sigma_{\ell\nu
				} \\
				& + \sigma_{jk}\sigma_{ki}\sum_{\mu\neq i,j,k,\ell}\sigma_{i\mu}\sigma_{\mu j} + \sigma_{jk}\sigma_{j\ell}\sum_{\mu\neq i,j,k,\ell}\sigma_{k\mu}\sigma_{\mu \ell}\\
				& +(\sigma_{jk}\sigma_{ki}\sigma_{j\ell})\sigma_{\ell i}.
			\end{aligned}
		\end{equation}
		The first three terms are all the terms in the Hamiltonian which contain $ \sigma_{jk} $, $ \sigma_{ki} $, and $ \sigma_{j\ell} $ separately, the next two terms contain $ \sigma_{jk}\sigma_{ki} $ and $ \sigma_{jk}\sigma_{j\ell} $, and the last term is a square that contains all three edges. We have
		\begin{equation}
			\langle\sigma_{jk}\sigma_{ki}\sigma_{j\ell}\rangle=\sum_{\{\sigma_{jk},\sigma_{ki},\sigma_{j\ell}=\pm 1\}}\sigma_{jk}\sigma_{ki}\sigma_{j\ell}\mathcal{P}(G).
		\end{equation}
		We can consider all the configurations of three edges in an open square and use mean-field approximation. We have
		\begin{equation}
			\begin{aligned}
				-\mathcal{H}^{+++}_{\sqcup}&=3(n-4)(n-5)\,o
				+2q(n-4)+p,\\
				-\mathcal{H}^{---}_{\sqcup}&=-3(n-4)(n-5)\,o
				+2q(n-4)-p,\\
				-\mathcal{H}^{-++}_{\sqcup}&=(n-4)(n-5)\,o
				-2q(n-4)-p,\\
				-\mathcal{H}^{+-+}_{\sqcup}&=(n-4)(n-5)\,o-p,\\
				-\mathcal{H}^{++-}_{\sqcup}&=(n-4)(n-5)\,o-p,\\
				-\mathcal{H}^{--+}_{\sqcup}&=-(n-4)(n-5)\,o+p,\\
				-\mathcal{H}^{-+-}_{\sqcup}&=-(n-4)(n-5)\,o+p,\\
				-\mathcal{H}^{+--}_{\sqcup}&=-(n-4)(n-5)\,o-2q(n-4)+p,
			\end{aligned}
		\end{equation}
		where  $ \mathcal{H}^{abc}_{\sqcup}\equiv\mathcal{H}_{\sqcup}(\sigma_{jk}=a,\sigma_{ki}=b, \sigma_{j\ell}=c) $. By defining $\Gamma(o)\equiv(n-4)(n-5)\,o$ and  $ \Sigma(o)\equiv(n-3)(n-4)\,o $, we can write: 
		\begin{widetext}
			\begin{equation}\label{opensquares}
				o=\frac{e^{3\beta\Gamma (o)+2\beta(n-4)q+\beta p} - e^{-3\beta\Gamma (o)+2\beta(n-4)q-\beta p} -e^{\beta\Gamma (o)-2\beta(n-4)q-\beta p} - 2\,e^{\beta\Gamma (o)-\beta p} + 2\,e^{-\beta\Gamma (o)+\beta p} +e^{-\beta\Gamma (o)-2\beta(n-4)q+\beta p}}
				{e^{3\beta\Gamma (o)+2\beta(n-4)q+\beta p} + e^{-3\beta\Gamma (o)+2\beta(n-4)q-\beta p} + e^{\beta\Gamma (o)-2\beta(n-4)q-\beta p} +  2\,e^{\beta\Gamma (o)-\beta p} + 2\,e^{-\beta\Gamma (o)+\beta p} +e^{-\beta\Gamma(o)-2\beta(n-4)q+\beta p}},
			\end{equation}
			\begin{equation}\label{triangles}
				r=\frac{e^{3\beta\Sigma (o)+3\beta(n-3)q} - e^{-3\beta\Sigma (o)+3\,\beta(n-3)q} -3\,e^{\beta\Sigma (o)-\beta(n-3)q} + 3\,e^{-\beta\Sigma (o)-\beta(n-3)q}}
				{e^{3\beta\Sigma (o)-3\,\beta(n-3)q} + e^{-3\beta\Sigma (o)-3\,\beta(n-3)q} + 3\,e^{\beta\Sigma (o)-\beta(n-3)q} + 3\,e^{-\beta\Sigma (o)-\beta(n-3)q}},
			\end{equation}
			\begin{equation}\label{squares}
				s=\frac{e^{4\beta\Gamma (o)+4\,\beta(n-4)\,q+\beta} + e^{-4\beta\Gamma (o)+4\beta(n-4)\,q+\beta} - 4\,e^{2\beta\Gamma (o)-\beta} - 4\,e^{-2\beta\Gamma (o)-\beta} + 6\,e^{\beta}}
				{e^{4\beta\Gamma (o)+4\beta(n-4)\,q+\beta} + e^{-4\beta\Gamma (o)+4\,\beta(n-4)\,q+\beta} + 4\,e^{2\beta\Gamma (o)-\beta} + 4\,e^{-2\beta\Gamma (o)-\beta} + 6\,e^{\beta}}.
			\end{equation}

		We have calculated $ p $ in Eq. (\ref{meanofedges}).
		For the mean value of triangles (\ref{triangles}) we used its Hamiltonian, which is
		\begin{equation}
			\begin{aligned}
				-\mathcal{H}_{_{\Delta}}&= \sigma_{ij}\sum_{\mu\neq i,j,k}\sum_{\nu\neq i,j,k}\sigma_{i\mu}\sigma_{j\nu}\sigma_{\mu\nu}
				+ \sigma_{jk}\sum_{\mu\neq i,j,k,\ell}\sum_{\nu\neq i,j,k,\ell}\sigma_{k\mu}\sigma_{j\nu}\sigma_{\nu\mu}
				+ \sigma_{ki}\sum_{\mu\neq i,j,k,\ell}\sum_{\nu\neq i,j,k,\ell}\sigma_{\mu k }\sigma_{\mu \nu}\sigma_{\nu i}\\
				&+\sigma_{ij}\sigma_{ki}\sum_{\ell\neq i,j,k}\sigma_{j\ell}\sigma_{\ell k}+\sigma_{ij}\sigma_{jk}\sum_{\ell\neq i,j,k}\sigma_{i\ell}\sigma_{\ell k}+\sigma_{ik}\sigma_{jk}\sum_{\ell\neq i,j,k}\sigma_{i\ell}\sigma_{\ell j}. \\
		\end{aligned}
		\end{equation}
		The mean value of squares (\ref{squares}) multiplied by minus one, represents the mean-field approximation for energy in our network and we derive it by using the following Hamiltonian:
		\begin{equation}
			\begin{aligned}
				-\mathcal{H}_{_{\square}}&=\sigma_{ki}\sum_{\mu\neq i,j,k,\ell}\sum_{\nu\neq i,j,k,\ell}\sigma_{i\mu}\sigma_{k\nu}\sigma_{\mu\nu}+ \sigma_{jk}\sum_{\mu\neq i,j,k,\ell}\sum_{\nu\neq i,j,k,\ell}\sigma_{\mu k }\sigma_{\mu \nu}\sigma_{\nu j}
				+\sigma_{j\ell}\sum_{\mu\neq i,j,k,\ell}\sum_{\nu\neq i,j,k,\ell}\sigma_{\mu j }\sigma_{\mu \nu}\sigma_{\nu \ell} \\
				&+  \sigma_{\ell i}\sum_{\mu\neq i,j,k,\ell}\sum_{\nu\neq i,j,k,\ell}\sigma_{\ell\mu}\sigma_{i\nu}\sigma_{\nu\mu}
				+\sigma_{jk}\sigma_{ki}\sum_{\mu\neq i,j,k,\ell}\sigma_{i\mu}\sigma_{\mu j} + \sigma_{jk}\sigma_{j\ell}\sum_{\mu\neq i,j,k,\ell}\sigma_{k\mu}\sigma_{\mu\ell}
				+ \sigma_{j\ell}\sigma_{\ell k}\sum_{\mu\neq i,j,k,\ell}\sigma_{k\mu}\sigma_{\mu j} \\
				&+ \sigma_{\ell i}\sigma_{ki}\sum_{\mu\neq i,j,k,\ell}\sigma_{\ell\mu}\sigma_{\mu k}
				+ \sigma_{jk}\sigma_{ki}\sigma_{j\ell}\sigma_{\ell i}.
			\end{aligned}
		\end{equation}
	\end{widetext}
	\end{appendices}

\end{document}